\documentclass[final, 3p, a4paper]{elsarticle}

\usepackage{papersettings}
\usepackage{mathsettings}
\usetikzlibrary{fit}
\usetikzlibrary{angles}
\usepackage{setspace}

\graphicspath{{./figures/}}

\journal{}

\newcommand{\gradient}{\nabla}

\begin{document}

\begin{frontmatter}
  \title{Eulerian framework for contact between solids represented as phase~fields}
  
    \address[ifbaddress]{Institute for Building Materials, ETH Zurich, Switzerland}
  \author[ifbaddress]{Flavio Lorez}
  \ead{florez@ethz.ch} 
  \author[ifbaddress]{Mohit Pundir}
  \author[ifbaddress]{David S. Kammer \corref{cor1}} \ead{dkammer@ethz.ch}
  \cortext[cor1]{Corresponding author}

\begin{abstract}
Mechanical contact between solids is almost exclusively modeled in Lagrangian frameworks. While these frameworks have been developed extensively and applied successfully to numerous contact problems, they generally require complex algorithms for contact detection and resolution. These challenges become particularly important when contact appears between solids with evolving boundaries, such as in systems where crystals grow in a constrained space. In this work, we introduce a fully Eulerian finite element framework for modeling contact between elastic solids tailored towards problems including evolving and intricate surfaces. The proposed approach uses a phase-field method that involves a diffuse representation of geometries on a fixed mesh, simplifying the modeling of evolving surfaces. Our methodology introduces a novel volumetric contact constraint based on penalty body forces, efficiently resolving the interpenetration of solids. We showcase the validity and versatility of our method through numerical examples, highlighting its ability to accurately capture complex solid-solid interactions. The Eulerian phase-field formulation greatly simplifies contact detection and its resolution. Furthermore, the framework can be straightforwardly coupled with other physical phenomena through the inclusion of multiple energy terms in the evolution of the phase-field. This enables multiphysics modeling, potentially providing a valuable tool for a wide range of applications involving chemically or physically evolving deformable solids in contact, as they commonly occur in deterioration processes of porous media. 
\end{abstract}


\end{frontmatter}


\section{Introduction}
\label{sec:introduction}

Contact between evolving solids plays a pivotal role in the analysis of growing objects in constraint space, encompassing phenomena like the emergence of corrosion precipitates in cementitious materials or the formation of ice crystals within porous rock. In such scenarios, accurately capturing contact interactions amid deformations and growth processes is very challenging for computational methods.
Solid-solid contact problems are almost exclusively modeled in a Lagrangian framework, where the discretization moves with the deforming objects. The very nature of a Lagrangian description, combining multiple reference systems, necessitates contact detection and discretization, which may impose challenges for large deformations or intricate interfaces~\cite{wriggers_computational_2006,Yastrebov2012,paggi2020computational,lengiewicz2011automation}.
Moreover, the Lagrangian framework presents difficulties in coupling contact with physical phenomena that are more naturally described in an Eulerian frame, such as evolving surfaces, fluid mechanics or phase transformations.
By adopting an Eulerian formulation, we aim to create a unified and holistic model capable of handling contact and its integration with a wider range of physical phenomena.

Contrary to the Lagrangian frame, in an Eulerian framework, the space is discretized, and objects are moved through it. 
In recent years, the fully Eulerian approach has gained popularity in the field of fluid-structure interaction (FSI)~\cite{dunne_eulerian_2006,richter_fully_2013,kamrin_reference_2012}.
Notably, some contributions in this domain have also incorporated the treatment of contact between submerged solids~\cite{valkov_eulerian_2015,teng_eulerian_2016,reder_phase-field_2021}. These studies have demonstrated the simplicity in handling collision within fully Eulerian simulations.
For solid-solid contact, however, research using a fully Eulerian framework has remained limited.
Few exceptions include Eulerian simulations of elastic solids in impact problems, during which one or both participating objects may be severely deformed~\cite{benson_multi-material_1995,benson_contact_2004}, and
a fully Eulerian simulator for colliding solids in the context of computer graphics~\cite{levin_eulerian_2011}. However, their formulation relies on specific, relatively complex numerical schemes to handle contact, making its adaption to other applications difficult.
Motivated by the advances in the context of FSI, we will show that a fully Eulerian framework has potential for an accurate and efficient treatment of solid-solid contact.

Traditionally, in the Eulerian framework, a single velocity field is employed to describe the motion of multiple bodies. This disallows relative displacement along the contacting surfaces between solids and, consequently, may introduce non-physical bonding between adjacent material~\cite{benson_mixture_1997}.
To sidestep this limitation entirely, we describe each solid body using its own set of Eulerian field variables. In essence, this is akin to a Lagrangian multi-body formulation, where each body is characterized by its unique reference system. This approach renders the different bodies independent by default but allows for coupling where necessary, facilitating the implementation of accurate and straightforward contact constraints.
A significant advantage of this approach compared to the Lagrangian formulation is the use of a single mesh for all bodies, which simplifies contact detection tremendously, as will be demonstrated in \cref{sec:contact-formulation}. 

In Eulerian frameworks, an additional solution field is used to capture the interface location. The most common methods are the volume of fluid~\cite{hirt_volume_1981}, the level-set~\cite{kamrin_reference_2012, osher_level_2003, sethian_level_1999} and the phase-field method~\cite{sun_sharp_2007,anderson_diffuse-interface_1998}. We will incorporate the phase-field method, where a continuous order parameter defines the domain, and the boundary is characterized by a diffuse hyperbolic transition layer of finite thickness. 
The phase-field method has been shown to be suitable for the purpose of implicit interface tracking~\cite{sun_sharp_2007}, and has emerged as a versatile tool for modeling various problems involving moving and evolving interfaces~\cite{du_chapter_2020}.
For instance, the method has been used for interface-capturing in modeling two-phase flow~\cite{aland_phase_2017,joshi_positivity_2018,mao_variational_2021,mao_interface_2023}, fracture propagation~\cite{bourdin2008variational,miehe2010thermodynamically,fei2020phase}, and fluid-structure interaction~\cite{mokbel_phase-field_2018,reder_phase-field_2021}.
One of the primary motivations for choosing the phase-field approach over the level-set method is its capability to model solidification within the same framework, which is achieved by evolving the phase-field under the influence of a carefully selected free-energy functional~\cite{tourret_phase-field_2022,boettinger_phase-field_2002}.
Additionally, advective distortion necessitates reinitialization of the level-set function, whereas the hyperbolic interface profile is naturally regularized~\cite{sun_sharp_2007,aland_phase_2017,mao_variational_2021}, as such contributing to the robustness and generality of the method. 
To model non-linear elastic continua, we employ the reference map technique proposed by Kamrin et al.~\cite{kamrin_reference_2012}.
In the Lagrangian framework, the spatial discretization is tied to the material points, making the deformation gradient readily available. However, in the Eulerian framework, where the mesh remains stationary, the deformation gradient is not intrinsically available. The reference map technique addresses this challenge by advecting a reference configuration, ensuring its availability throughout the analysis, as it has been successfully demonstrated for FSI within a fully Eulerian framework~\cite{dunne_eulerian_2006,kamrin_reference_2012,valkov_eulerian_2015,jain_conservative_2019}. The approach has also been named characteristic mapping method in the more general context of advection~\cite{mercier_characteristic_2020}.

The paper is organized as follows.
In \cref{sec:method}, we introduce the novel Eulerian-frame continuum formulation for elastic solids. Next, in \cref{sec:contact-formulation}, we present the proposed volumetric contact constraint, enabling the coupling and resolution of contact between bodies defined on a single fixed mesh.
The discretization and numerical implementation details are provided in \cref{sec:numerical_implementation}.
We then demonstrate the validity and versatility of our method through four numerical examples in \cref{sec:examples}.
Finally, in \cref{sec:discussion,sec:conclusion}, we discuss the implications of our method and draw a conclusion.


\section{Eulerian continuum phase-field formulation}
\label{sec:method}

\begin{figure}
    \centering
    \includegraphics{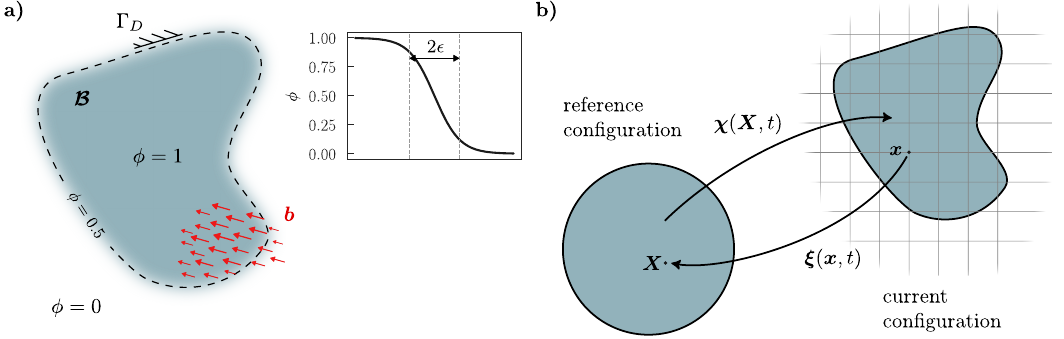}
    \caption{
    \textbf{(a)} Illustration of a diffuse body $\mathcal{B}$. Its domain is implicitly defined by regions where $\phi>0.5$. The body may be subject to Dirichlet BCs and a force field $\vector{b}$. The diffuse boundary, where $\phi$ smoothly transitions from $1$ to $0$, is shown in 1D. The parameter $\epsilon$ controls the width of the diffuse interface.
    \textbf{(b)} Illustration of the reference map $\vector{\xi}$. $\vector{\chi}(\vector{X}, t)$ is the motion function mapping the material points to $\x$. $\vector{\xi}(\x, t)$ is the inverse motion function mapping the Eulerian spatial coordinates $\x$ to the material points defined in the reference configuration.
    }
    \label{fig:eulerian_motion}
\end{figure}



Consider a computational domain in $\mathrm{D}$ dimensions $\Omega\subset\mathbb{R}^ \mathrm{D}$ in which we place elastic continua. We denote $\mathcal{B}$ as a deformable body subject to an external force field $\vector{b}$ and Dirichlet boundary conditions on $\Gamma_D$.
The solid $\mathcal{B}$ is implicitly represented by a phase-field variable $\phi: \Omega\times[0,T]\rightarrow\mathbb{R}$ such that $\phi$ takes the value 1 inside the deformable body $\mathcal{B}$ and 0 elsewhere (see \cref{fig:eulerian_motion}). 

The phase-field method involves a diffuse representation of the boundary $\partial\mathcal{B}$, characterized by
the minimization of the free energy functional~\cite{cahn_free_1958},
\begin{equation}\label{eq:free-energy-functional}
    \mathcal{E}_{\text{PF}}(\phi) = \int_\Omega \left( \dfrac{\epsilon^2}{2}(\nabla\phi)^2 + g(\phi) \right) \mathrm{d\Omega} ~, 
\end{equation}
where $g(\phi)=\phi^2(1-\phi)^2$ represents a double well potential with its minima located at $0$ and $1$, and $\epsilon^2$ is the energy gradient coefficient. 
The variational derivative of \cref{eq:free-energy-functional} with respect to $\phi$ is called the chemical potential and is given by
\begin{equation}
    \dfrac{\delta\mathcal{E}_{\text{PF}}}{\delta\phi} = -\epsilon^2\nabla^2\phi + g'(\phi) ~.
\end{equation}
Finding the root $\delta\mathcal{E}/\delta\phi=0$ reveals the well-known equilibrium interface profile, as shown in \cref{fig:eulerian_motion}{a},
\begin{equation}\label{eq:interface_profile}
    \phi(r) = \dfrac{1}{2}\left[ 1 - \tanh{\left(\dfrac{r}{\sqrt{2}\epsilon}\right)} \right] ~,
\end{equation}
where $r$ is the signed distance from the sharp interface.

If $\mathcal{B}$ is deformed by a displacement field $\uu$, its phase-field $\phi$ must be evolved accordingly to capture the deformed shape. We start by observing that the phase of some followed particle should not change. Hence, the material time derivative of the phase-field is equal to $0$,
\begin{equation}\label{eq:material_time_phi}
    \dfrac{D\phi}{Dt} = \dfrac{\partial\phi}{\partial t} + \dfrac{\partial \uu}{\partial t} \cdot \nabla\phi = 0 ~.
\end{equation}
When \cref{eq:material_time_phi} is used directly, $\phi$ is unconstrained. Given the arbitrary nature of $\uu$, this may result in $\phi$ obtaining values outside the desired range $\phi\in[0,1]$. In addition, the interface may be squeezed or stretched, leading to a deviation from the desired interface thickness. This has been referred to as advective distortion~\cite{sun_sharp_2007,mao_variational_2021} and may result in numerical issues.
To limit advective distortion, we consider the gradient flow towards minimizing the diffuse interface energy using the Cahn-Hilliard equation~\cite{cahn_free_1958,li_solving_2009,aland_phase_2017}:
\begin{equation}\label{eq:system-phi}
    \dfrac{\partial\phi}{\partial t} + \dfrac{\partial \uu}{\partial t}\cdot\nabla\phi = \nabla\cdot\left(-\mathcal{M}~\nabla\left(\dfrac{\delta\mathcal{E}_{\text{PF}}}{\delta\phi}\right) \right) ~,
\end{equation}
where $\mathcal{M}$ is the mobility coefficient. The left-hand side describes the displacement of $\phi$ through linear advection subject to the background velocity $\partial\uu / \partial t$, and the right-hand side is responsible for a diffusive relaxation towards the equilibrium profile \cref{eq:interface_profile}. To this end, the main purpose of the phase-field is the implicit representation of the body's boundary. Consequently, we are primarily interested in pure advection of the interface rather than minimization of its interfacial energy. The inclusion of the chemical potential can, therefore, be seen as a regularization measure to a standard linear advection to avoid advective distortion of the phase-field's diffuse interface.
Finally, we note that the mobility parameter $\mathcal{M}$ controls the magnitude of regularization.

To capture the deformation state of $\mathcal{B}$, we employ the reference map technique (RMT). For a detailed description of the method, the reader is referred to~\cite{kamrin_reference_2012}. Thus, we introduce the reference map $\vector{\xi}:\Omega\times[0,T]\rightarrow\mathbb{R}^\mathrm{D}$, an Eulerian field, as the inverse motion function (see \cref{fig:eulerian_motion}{b}), \ie
\begin{equation}
    \vector{\xi}(\x, t) = \vector{\chi}^{-1}(\vector{X}, t) = \vector{X}(\x, t), \quad \vector{\xi}(\x, 0)=\x ~.
\end{equation}
It can be understood as a vector field pointing towards the location of a material point currently at a position $\x$ in the reference configuration $\vector{X}$. The mapping $\vector{\chi}(\vector{X}, t)$ is known as the motion function.
Mathematically, it is known that the reference map constitutes a diffeomorphism~\cite{mercier_characteristic_2020}.
Considering that,
\begin{equation}
    \label{eq:rmt_defgradient}
    \tensor{F}(\vector{x}, t) = \dfrac{\partial\x}{\partial\vector{X}} = \left(\dfrac{\partial\vector{X}}{\partial\x}\right)^{-1} = \left(\gradient\vector{\xi}(\vector{x}, t)\right)^{-1} ~, 
\end{equation}
it is clear that the reference map carries sufficient information to retrieve the deformation gradient at all times. Therefore, the RMT allows modeling any history-independent material where the response is a function of the current deformation only.
Observing that for a unique tracer particle, the reference position is constant, $\vector{\xi}$ is evolved by linear advection~\cite{kamrin_reference_2012}, \ie
\begin{equation}\label{eq:system-zeta}
    \dfrac{\partial \vector{\xi}}{\partial t} + \dfrac{\partial \uu}{\partial t}\cdot \gradient \vector{\xi} = \vector{0} ~.
\end{equation}

While the phase-field $\phi$ captures the interface, the reference map $\vector{\xi}$ captures the deformation of the body. Hence, the combination of $\phi$ and $\vector{\xi}$ is sufficient to fully describe the configuration of the solid $\mathcal{B}$ at time $t$. 
In this work, we adopt a static implicit scheme. 
This means, given the current configuration $\phi_i$ and $\vector{\xi}_i$, the force field $\vector{b}$ and Dirichlet boundary conditions on $\Gamma_D$ (see \cref{fig:eulerian_motion}{a}), finding an updated configuration $\phi_{i+1}$ and $\vector{\xi}_{i+1}$ in equilibrium.
The body is in static equilibrium if
\begin{gather}
    \gradient \cdot \tensor{\sigma} - \vector{b} = 0 , \label{eq:system-equilibrium} \\
    \tensor{\sigma} = \phi \cdot \hat{\tensor{\sigma}}(\tensor{F}), \quad
    \tensor{F} = \tensor{F}(\vector{\xi}), \quad
    \uu = \Bar{\uu} \in \Gamma_D , \notag
\end{gather}
where $\hat{\tensor{\sigma}}(\tensor{F})$ is the elastic constitutive material response and $\Bar{\uu}$ is the prescribed displacement.

In order to find the updated configuration at $t=i+1$, we introduce a body-specific displacement increment field $\Delta\uu : \Omega\times[0, T]\rightarrow \mathbb{R}^\mathrm{D}$ to the solution space, which advects the current to the updated configuration in equilibrium. Crucially, for a multi-body system, each body receives its own displacement increment field to describe its motion.
\cref{eq:system-equilibrium,eq:system-phi,eq:system-zeta} are a closed Eulerian system to compute the equilibrated configuration of $\mathcal{B}$.


\section{Phase-field volumetric contact constraint}
\label{sec:contact-formulation}
In this section, we present a method to incorporate frictionless contact using penalty-based body forces.
We consider two bodies $\mathcal{B}_k$ and $\mathcal{B}_l$.
Following the formulation in \cref{sec:method}, each body $\mathcal{B}_k$ and $\mathcal{B}_l$ evolves independently, characterized by its own set of field variables in the Eulerian framework. To enforce non-penetration between the two bodies, we introduce volumetric penalty forces that couple them together.
Our novel approach focuses on minimizing the intersection between their phase-fields as a way to resolve contact. By introducing a volumetric body force that acts only within the intersection volume, we effectively target the regions of contact without constraining the deformations directly.

\begin{figure}
    \centering
    \includegraphics{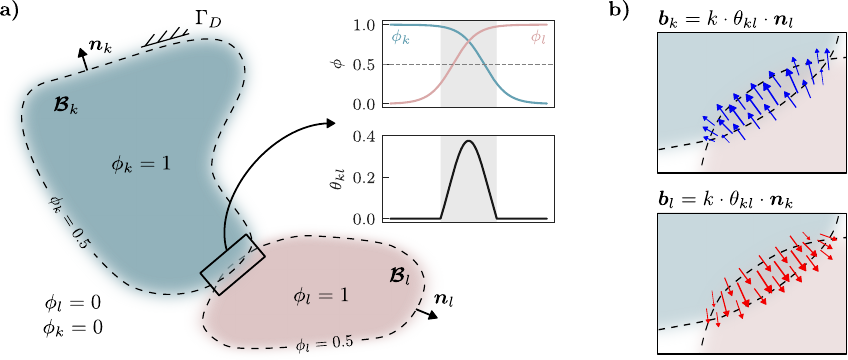}
    \caption{Schematic of the contact constraint. \textbf{(a)} The two phase-fields $\phi_k$ and $\phi_l$ represent the domains of the bodies. Either body may be subject to Dirichlet BCs. The intersection in the contact zone is illustrated in a 1D cut normal to the contact interface, showing the general shape of $\theta$.
    \textbf{(b)} Body forces of the contact constraint acting on the blue and red body, respectively. The force field acts normal to the gradient of the other body's phase-field. The intersection and forces are exaggerated for illustration purposes.  
    }
    \label{fig:contact-introduction}
\end{figure}

We regard two bodies to be in contact if their $\phi = 0.5$ isosurface overlap.
Thus, we define the scalar intersection variable of two bodies $\mathcal{B}_k$ and $\mathcal{B}_l$ as
\begin{equation}
    \label{eq:theta}
    \theta_{kl} := \left\langle \phi_k\cdot\phi_l - \dfrac{1}{4} \right\rangle^+ ~,
\end{equation}
where $\langle \cdot \rangle^+$ is the Macaulay bracket. Due to the symmetry in the interface profile of the phase-fields, $\theta_{kl}$ will be non-zero only in the immediate proximity of overlapping zones.
Although bounded by the upper limit, $\theta_{kl}=3/4$, the intersection measure is dependent on the penetration depth, ensuring that the contact force, $\vert\boldsymbol{b}_k\vert = f(\theta_{kl})$, increases with increasing penetration. 
The fact that collisions should be detected if the multiplication of two phase-fields $\phi\in[0,1]$ exceeds $1/4$ was also acknowledged by Reder~et~al.~\cite{reder_phase-field_2021}.

The penalty force field shall be aligned normal to the other body's diffuse boundary, which is given by the normalized gradient of the phase-field,
\begin{equation}\label{eq:body_forces}
    \vector{n}_l = \dfrac{\nabla\phi_l}{\vert\nabla\phi_l\vert} ~.
\end{equation}

Given the intersection metric $\theta_{kl}$ and the normals $\boldsymbol{n}_k$ and $\boldsymbol{n}_l$, any appropriate function for the contact forces, $\boldsymbol{b}_k = f(\theta_{kl}, \boldsymbol{n}_l)$, can be constructed.
For frictionless contact, a constant penalty parameter $k$ and a linear combination of \cref{eq:theta,eq:body_forces} is well suited. Thus, in this paper, we define the body forces originating from contact on either body as
\begin{equation}
    \vector{b}_k = k\cdot\vector{n}_l\cdot\theta_{kl}, \quad \vector{b}_l = k\cdot\vector{n}_k\cdot\theta_{kl} ~.
\end{equation}
These forces are included in \cref{eq:system-equilibrium} to find the updated configurations for both bodies considering contact.
A key advantage here is the simplicity of detecting the intersection $\theta_{kl}$ between bodies $\mathcal{B}_k$ and $\mathcal{B}_l$ using a simple function of the solution space. This enables us to solve the system with contact implicitly and monolithically, without the need of specifying any constraints. 
While this penalty approach does not entirely resolve penetration, leading to slight differences between the normals of the two bodies, the contact forces are practically in equilibrium for suitable values of the penalty parameter $k$.


\section{Numerical implementation}
\label{sec:numerical_implementation}
The Eulerian system presented in \cref{sec:method}, in combination with the contact constraint presented in \cref{sec:contact-formulation}, describes a non-linear system which can be numerically discretized. 
\Cref{eq:system-phi,eq:system-zeta} are given in rate-form and must be integrated in time to obtain a discrete form that can be implicitly solved for the equilibrium configuration.
By employing the Crank-Nicolson implicit time integration scheme and considering a pseudo time step $\Delta t=1$, \cref{eq:system-phi,eq:system-zeta} become
\begin{gather}
    (\phi_{i+1} - \phi_i) + \Delta\uu \cdot  \nabla\left(\dfrac{\phi_{i+1}+\phi_{i}}{2}\right) + \nabla\cdot\mathcal{M} \left[\nabla(g'(\phi_{i+1}) - \epsilon^2\nabla^2\phi_{i+1}) \right] = 0 ~, \label{eq:phi-discrete}\\
    (\vector{\xi}_{i+1} - \vector{\xi}_i) + \Delta\uu^T \cdot \nabla\left(\dfrac{\vector{\xi}_{i+1}+\vector{\xi}_i}{2} \right) = \vector{0} ~. \label{eq:zeta-discrete}
\end{gather}
The system is completed by the linear momentum equation (\cref{eq:system-equilibrium}).

We use FEniCS~\cite{logg_dolfin_2010} to employ the finite element method to monolithically solve the system given by \cref{eq:system-equilibrium,eq:phi-discrete,eq:zeta-discrete}.
By multiplying the strong form with the corresponding test function $\delta f \in V_\delta = \mathcal{H}_0^1(\Omega):=\{\delta f\in \mathcal{L}^2 | \delta f=0 \text{ on } \Gamma_D  \}$ and using integration by parts, we obtain the variational form~\cite{belytschko_nonlinear_2014}.

The weak form of the equilibrium equation \cref{eq:system-equilibrium} reads
\begin{equation}
    \label{eq:variational_1}
    \int_\Omega \phi  \cdot \hat{\tensor{\sigma}} \colon \gradient\delta\Delta\uu  \mathrm{d\Omega}
        - \int_\Omega \vector{b} \cdot \delta\Delta\uu ~\mathrm{d\Omega} = 0 ~.
\end{equation}

The advective Cahn-Hilliard equation \cref{eq:phi-discrete} is a fourth-order equation. Casting it in a weak form would result in second-order derivatives, and first-order Lagrangian elements would not be sufficient. Therefore, we rephrase the equation using two coupled second-order equations:
\begin{align}
    \label{eq:variational_2}
    \int_\Omega (\phi_{i+1} - \phi_{i}) \delta\phi ~\mathrm{d\Omega}
        + \int_\Omega  \Delta\uu \cdot \nabla\left(\dfrac{\phi_{i+1}+\phi_i}{2}\right) \delta\phi ~\mathrm{d\Omega}
        + \mathcal{M} \int_\Omega \left(\nabla\mu\cdot\nabla\delta\phi\right) ~\mathrm{d\Omega} &= 0 ~, \\
    \label{eq:variational_3}
    \int_\Omega \mu \delta\mu ~\mathrm{d\Omega}
        - \int_\Omega \epsilon^2 \left(\nabla\phi \cdot \nabla\delta\mu\right) ~\mathrm{d\Omega}
        - \int_\Omega g'(\phi) \delta\mu ~\mathrm{d\Omega} &= 0 ~.
\end{align}

The evolution of the reference map \cref{eq:zeta-discrete} in weak form reads
\begin{equation}
    \label{eq:variational_4}
    \int_\Omega (\vector{\xi}_{i+1} - \vector{\xi}_{i}) \cdot \delta\vector{\xi} ~\mathrm{d\Omega}
        + \int_\Omega \Delta\uu^T\cdot \nabla \left(\dfrac{\vector{\xi}_{i+1}+\vector{\xi}_i}{2} \right) \cdot \delta\vector{\xi} ~\mathrm{d\Omega} = 0 ~.
\end{equation}

Then, the variational problem reads: Find the functions $(\Delta\uu, \vector{\xi}, \phi, \mu) \in V = \mathcal{H}^1(\Omega)$ for every body such that \cref{eq:variational_1,eq:variational_2,eq:variational_3,eq:variational_4} are satisfied.
The variational formulation is solved monolithically, \ie fully coupled. The mixed function space contains $2\mathrm{D}+2$ orthogonal basis functions per body. It shall be mentioned here, given the linearity of \cref{eq:zeta-discrete}, that in many cases $\Delta\uu$ can be expressed in terms of $\boldsymbol{\xi}$ and thus the solution space is reduced. Schematically, the following system is solved:
\begin{equation}
    \underbrace{
        \begin{bmatrix}
            \mathbf{K_u} & \mathbf{K_{u\xi}} & \mathbf{K_{u\phi}} & 0 \\
             & \mathbf{K_\xi} & 0 & 0 \\
             & & \mathrm{K_\phi} & \mathrm{K_{\phi\mu}} \\
             & & & \mathrm{K_\mu}
        \end{bmatrix}
    }_{\mathbf{K^{(k)}}}
    \cdot
    \underbrace{
        \begin{bmatrix}
            \Delta\uu \\ \vector{\xi} \\ \phi \\ \mu
        \end{bmatrix}
    }_{\mathbf{u^{(k)}}}
    - \underbrace{\begin{bmatrix}
        \vector{b}(\phi) \\ \vector{0} \\ 0 \\ 0
    \end{bmatrix}}_\mathbf{f^{(k)}}
    = \vector{0}
\end{equation}

Considering a system with two bodies $\mathcal{B}_k$ and $\mathcal{B}_l$, the coupling between them is achieved solely through the force vectors $\mathbf{f^{(k)}}(\phi_k, \phi_l)$. Hence, the combined monolithic problem reads:
\begin{equation}
    \begin{bmatrix}
        [\mathbf{K^{(k)}}] &  0 \\  & [\mathbf{K^{(l)}}]
    \end{bmatrix}
    \cdot
    \begin{bmatrix}
        \mathbf{u^{(k)}} \\ \mathbf{u^{(l)}}
    \end{bmatrix}
    - \begin{bmatrix}
        \mathbf{f^{(k)}} \\
        \mathbf{f^{(l)}}
    \end{bmatrix}
    = \vector{0}
\end{equation}

\section{Examples}
\label{sec:examples}

In this section, we present numerical results obtained using the proposed method. Throughout this paper, we make use of dimensionless units. We begin by demonstrating that the method yields a traction profile consistent with the analytical Hertz solution for contact under infinitesimal displacement conditions. This validation step ensures the accuracy and reliability of the method in capturing basic contact behaviors.

Next, we investigate the method's performance in handling large deformations involving Neo-Hookean material. By subjecting the material to significant strains, we assess the method's ability to accurately capture nonlinear material responses and validate its applicability in modeling more realistic material behaviors. Furthermore, we showcase the method's versatility by applying it to more complex domains. Despite the increased complexity of these domains, we demonstrate that the method's implementation remains efficient and requires minimal additional effort. This capability opens the door to simulating complex systems and structures with varying geometries, enabling a wider range of applications for the proposed numerical approach.

Overall, these numerical results validate the accuracy, robustness, and versatility of our method, making it a promising tool for simulating various contact and deformation scenarios, including those involving nonlinear material behavior and complex domains.

\subsection{Hertz contact problem on a rigid plane}
\label{subsec:hertz_rigid}

\begin{figure}[t]
    \centering
    \includegraphics{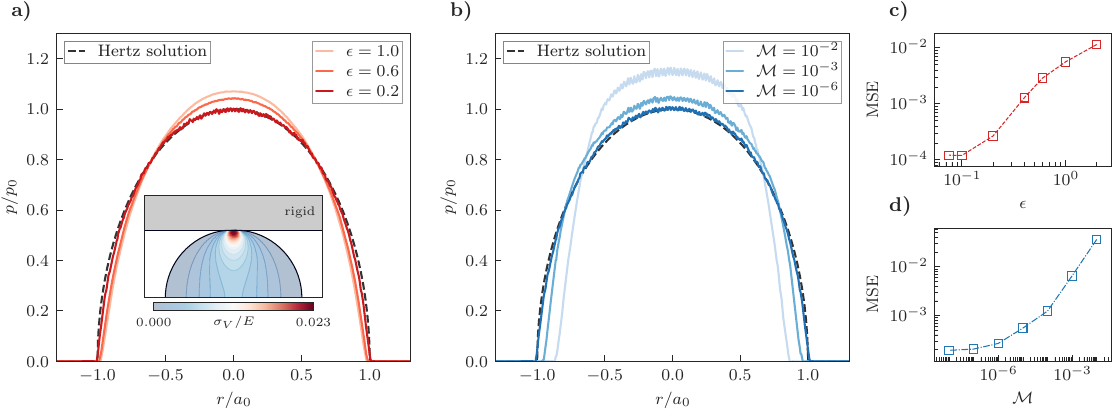}
    \caption{Analysis of the traction profile for different values for $\epsilon$ and $\mathcal{M}$ for contact between a deformable cylinder and a rigid plane.
    \textbf{(a)} Normalized traction profile for varying width of the diffuse interfaces. The inset shows the simulation setup and the Von Mises stress for $\epsilon=0.2$.
    \textbf{(b)} Normalized traction profile for varying mobility of the Cahn-Hilliard equation (for $\epsilon=0.2$).
    \textbf{(c, d)} Mean squared error (MSE) of the obtained traction profile compared to the Hertz profile for increasing $\epsilon$ and $\mathcal{M}$, respectively. }
    \label{fig:hertz_rigid}
\end{figure}

We consider contact between a deformable cylinder with radius $R=10$ and a rigid plane in 2D (see inset in \cref{fig:hertz_rigid}{a}). The rigid plane is progressively moved downwards to compress the cylinder by a maximum of 1\% of its radius. We model the cylinder as a linear elastic Saint Venant-Kirchoff material~\cite{belytschko_nonlinear_2014}. The constitutive law is characterized by $\hat{\tensor{\sigma}} = \lambda\mathrm{tr}(\tensor{E})I + 2\mu\tensor{E}$, where $\lambda, \mu$ are the Lamé constants and $\tensor{E}$ is the Green strain.
We use a dimensionless setup, a Young's modulus of $E=1$ and a Poisson's ratio of $\nu=0.2$.
The rigid plane is introduced to the domain by a constant phase-field $\phi_{\text{rigid}}$ which is used to compute the intersection $\theta$ equivalent to the case with two deformable bodies.

The Hertz solution is characterized by the contact zone width and the maximum pressure~\cite{johnson_contact_1987}
\begin{equation}\label{eq:hertz_solution}
    a_0 = \left(\dfrac{4PR}{\pi E}\right)^{1/2} ~, \quad
    p_0 = \dfrac{2P}{\pi a_0} ~,
\end{equation}
where $P$ is the total contact force. The elliptic traction profile is then given by
\begin{equation}
    p_{\text{Hertz}}(r) = \dfrac{2P\sqrt{a^2-r^2}}{\pi a^2} ~.
\end{equation}

In our model, we compute the traction profile by integrating the contact body force $\vector{b}$ normal to the interface, which implies, here,  $p(x)=\int_y b_y ~\mathrm{dy}$. The total contact force is the volume integral, \ie $P=\int_\Omega \vector{b}\cdot\vector{n} ~\mathrm{d\Omega}$.
\Cref{fig:hertz_rigid} demonstrates the excellent agreement between the traction profile obtained from our numerical simulations and the analytical Hertz profile. Specifically, we investigate the impact of the two primary model parameters, $\epsilon$ and $\mathcal{M}$.

In \cref{fig:hertz_rigid}{a/c}, we observe the impact of reducing the diffuse interface width controlled by the parameter $\epsilon$ while keeping $\mathcal{M}=10^{-6}$ constant. As expected, decreasing $\epsilon$ results in improved agreement with the analytical solution. We note that for a constant value of the mobility parameter, if $\epsilon$ were decreased further, the deviation from the analytical solution would increase after some point. This is due to the heightened gradient $\nabla\phi$, which amplifies the gradient flow effect in \cref{eq:system-phi} as also observed in previous work~\cite{aland_phase_2017}, which found that $\mathcal{M}$ should be scaled proportionally to $\epsilon$. 


\Cref{fig:hertz_rigid}{b/d} illustrates the effect of the mobility parameter $\mathcal{M}$. We note that the result accuracy suffers for large values of $\mathcal{M}$ as a consequence of the increased influence of the diffuse surface energy. For large values of $\mathcal{M}$, the compressed cylinder tends to maintain a circular shape to reduce its surface energy, which is a deviation from the expected result.
This observation highlights the deliberate trade-off between introducing an approximation error by considering the Cahn-Hilliard equation and $\mathcal{M}$ while gaining the benefit of limiting advective distortion. While this effect may not be apparent for small deformations, in fact, even $\mathcal{M}=0$ results in no severe distortion, it becomes crucial for accurately modeling large deformations and motions. 

For all shown cases, we use a finely discretized mesh to ensure that enough finite elements are present within the contacting region and across the diffused boundary such that the results are mesh-independent. Therefore, lower values of $\epsilon$ necessitate finer meshes to maintain precision. Too coarse meshes cause discrepancies to the traction profile, particularly in the form of oscillations, but the width of the contact zone remains nearly unaffected.

\subsection{Hertz contact problem on a deformable plane}
\label{subsec:hertz_deformable}

\begin{figure}[tb]
    \centering
    \includegraphics{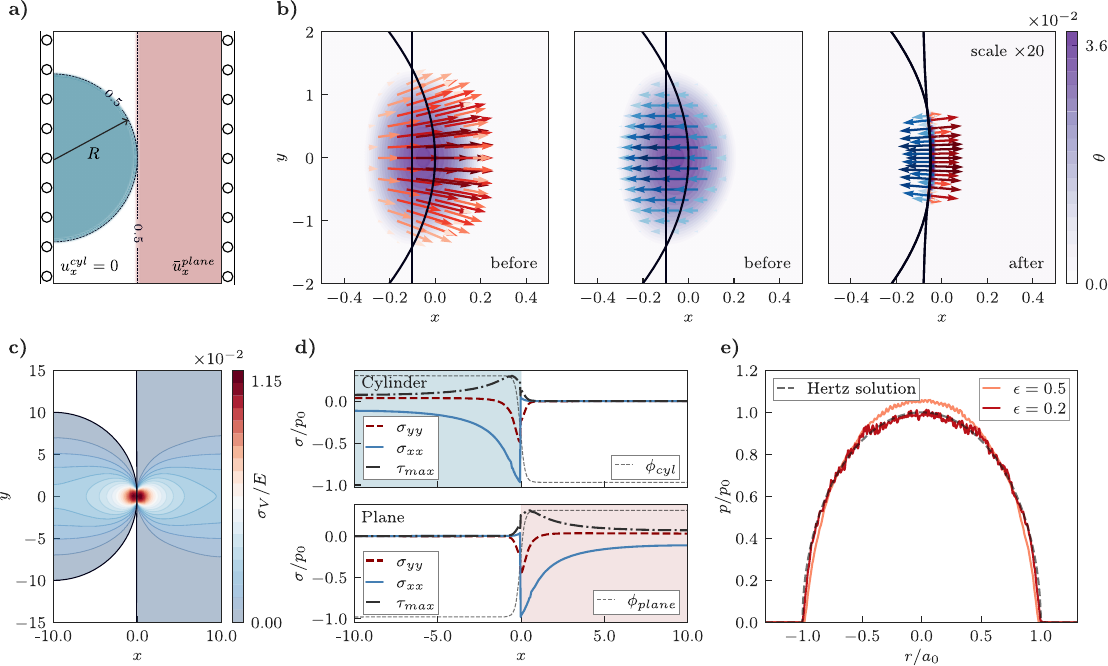}
    \caption{Numerical example of contact between a deformable cylinder and a deformable plane.
    Displacement controlled simulation ($u_{x, \text{plane}}=r/100$) with a deformable rigid and a deformable plane ($\epsilon = 0.5$, $\mathcal{M}~=~10^{-6}$).
    \textbf{(a)} Simulation setup where the two phase-fields are shown in blue and red. The horizontal displacement is fixed for the cylinder and the plane on the left and right boundary, respectively.
    \textbf{(b)} Body forces of the phase-field volumetric contact constraint. First, when the plane has been advected, but both bodies remain unstressed (before equilibrating). Red and blue arrows show the force field acting on the plane and on the cylinder, respectively. Lastly, the forces in the final equilibrated configuration. The $x$-axis is stretched for visualization purposes. 
    \textbf{(c)} Von Mises stresses in the system. Data is filtered to only show in the domain where $\phi>0.5$.
    \textbf{(d)} Bulk stresses in the cylinder (top) and in the plane (bottom) for the cross-section at $y=0$. Stresses are normalized with the peak contact pressure.
    \textbf{(e)} Normalized traction profile compared with the analytical Hertz solution.}
    \label{fig:hertz_deformable}
\end{figure}

Next, we apply the proposed methodology to simulate contact between two deformable bodies subject to small deformations. 
To this end, we consider a deformable cylinder of radius $R=10$ that is compressed by a deformable plane (see \cref{fig:hertz_deformable}{a}). We prescribe the plane's horizontal displacement at the right domain boundary $\Bar{u}_x = R/100$. 

The contact body forces are influenced by both the intersection $\theta$ and the normals of the phase-fields, as shown in \cref{eq:body_forces}. 
\cref{fig:hertz_deformable}{b} shows the force fields before and after enforcing the contact constraint. In the first two figures, we present the force fields for the plane and the cylinder, respectively, for a potential initial guess. The plane has moved into the cylinder, but the contact constraint has not yet exerted any influence. Hence, both bodies are stress-free and \cref{eq:system-equilibrium} is clearly not satisfied. Notably, the direction of the force field acting on the plane is normal to the gradient of the cylinder, and vice versa.
The third figure in \cref{fig:hertz_deformable}{b} illustrates the forces in the equilibrated state. It is important to note that, due to the penalty approach used in our method, the intersection between the cylinder and the plane is not entirely removed. However, it can be reduced by choosing an appropriate value for $k$.

Visualizing the stress distribution plays a crucial role in validating the reliability of our results. Leveraging the reference map, we compute the Von Mises stresses and illustrate them in \cref{fig:hertz_deformable}{c}. As expected, the stresses are highest in the immediate proximity of the contact region.
\Cref{fig:hertz_deformable}{d} further confirms these results, as it demonstrates that the normal stress is fully transmitted from cylinder to plane across the contact interface. We note that the maximal principal shear stress $\tau_{max}$ is observed inside the body and not at the surface. This is in agreement with the analytical solution obtained by Hertz~\cite{johnson_contact_1987}. 
Finally, we examine the traction profile at the contacting interface and compare it to the Hertz solution (\cref{eq:hertz_solution}). \Cref{fig:hertz_deformable}{e} displays the traction profile, and it shows results similar to those in \cref{fig:hertz_rigid}{a}. We note that minute discrepancies from the analytical solution are not necessarily of methodological nature, but may also arise from the finite size of our computational domain, in contrast to the idealized infinite domain assumed for the analytical derivation of the Hertz solution.  

The excellent agreement between our numerical results and the analytical solution validates the accuracy of our proposed method for contact between two deformable bodies. By analyzing the stress distribution, we gain insights into the contact mechanics, and the capability of the reference map technique to accurately capture the stress states in elastic bodies.

\subsection{Large deformations}
\label{subsec:large_deformations}

\begin{figure}[tb]
    \centering
    \includegraphics{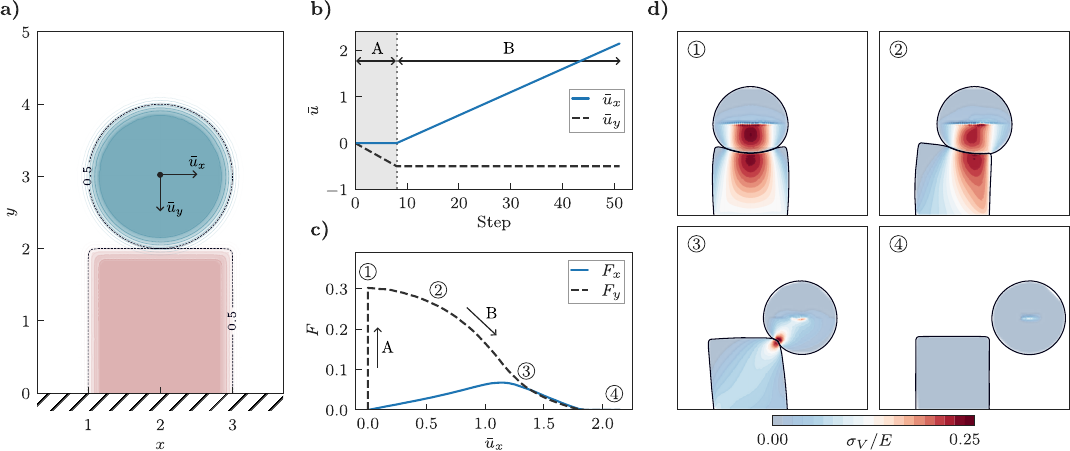}
    \caption{Displacement controlled disk compressing a stationary box, showing the capacity to handle large deformations with both geometrical and material nonlinearity.
    \textbf{(a)} Simulation setup, where the two phase-fields are shown in blue and red. The disk's movement is prescribed by $\vector{\Bar{u}}$ and the bottom boundary is fixed for the box.
    \textbf{(b)} The prescribed displacement. In Phase A, the disk is moved vertically by $-0.5$. In Phase B, the disk is moved horizontally until the end of the analysis.
    \textbf{(c)} Total horizontal and vertical traction forces over the course of the analysis. 
    \textbf{(d)} Von Mises stress at snapshots at the steps indicated in (c).    }
    \label{fig:example_box}
\end{figure}

To demonstrate our method's capability to model large deformations, we consider a circular disk with a radius of $R=1$ compressing a rectangular box with dimensions $2\times 2$, as illustrated in \cref{fig:example_box}{a}. The computational domain is discretized into a structured grid, where the two phase-fields $\phi_1$ and $\phi_2$ are initialized to describe the disk and the box's respective domains.
We prescribe the movement of the disk center, denoted by $\vector{\Bar{u}}$ and shown in \cref{fig:example_box}{b}. Initially, in phase A, the disk is moved downwards by a displacement of $0.5$. Subsequently, in phase B, the disk is moved horizontally until the end of the analysis. The vertical displacement $\Bar{u}_y$ is prescribed along the entire center axis, while the horizontal displacement $\Bar{u}_x$ is constrained only near the center, up to $0.1$ away from the disk center.
Throughout the simulation, the box displacement remains fixed at the bottom, providing a stable boundary condition.

To cope with large deformations, we employ the Neo-Hookean hyper-elastic material law for both bodies, where the stress-strain relationship is given by
\begin{equation}
    \hat{\tensor{\sigma}} = \lambda\ln{J} \tensor{C}^{-1} + \mu\left(\tensor{I} - \tensor{C}^{-1}\right) , \quad
    \tensor{C} = \tensor{F}^T\cdot \tensor{F} ,
\end{equation}
where $\tensor{C}$ is the right Cauchy-Green tensor, $J=\det\tensor{F}$ and $\lambda, \mu$ are the Lamé constants.

\Cref{fig:example_box}{c} shows the total contact force components $F_x$ and $F_y$ over the course of the analysis. In combination with \cref{fig:example_box}{d}, this figure effectively illustrates the modeled problem's behavior during the simulation. 
In phase A, the problem exhibits symmetry, resulting in vertical forces only, as shown in \cref{fig:example_box}{c}.
In phase B, the asymmetry introduced by the disk's horizontal movement leads to horizontal forces, even in the absence of friction. Consequently, the box experiences a tilting motion due to the asymmetrical disk position (see \cref{fig:example_box}{d}).
As the disk moves past the box, the contact forces gradually decrease and eventually collapse to zero again, signifying the end of the contact between the disk and the box.

Due to the lack of a theoretical or numerical reference solution, we qualitatively verify our results on several key observations. First and foremost, the original shapes of the box and the disk are accurately retrieved. The final configuration is practically unstressed, apart from the disk center, which maintains small residual stresses introduced by over-constraining the degrees of freedom when prescribing its displacement. However, these small residual stresses do not significantly affect the overall stress distribution.
Moreover, the stress distribution is perfectly symmetrical when it should be, and it becomes asymmetric in a reasonable manner when expected. We find no indication that the obtained stress distributions are unnatural for the considered system.

Aside from mechanical results, this example shows the necessity for the advective Cahn-Hilliard equation (see \cref{eq:system-phi}) in the governing equation for the phase-field.
Large linear advection of the phase-field causes advective distortion in the diffuse interface profile~\cite{mao_variational_2021, mao_interface_2023}. Here, this is especially critical for the disk because it is advected along a relatively long path. 
The mobility coefficient, here set to $\mathcal{M} = 10^{-7}$, controls the influence of the Cahn-Hilliard term, which naturally regularizes the hyperbolic tangent profile but introduces undesired interface displacements, especially for large curvature~\cite{sun_sharp_2007}.
When $\mathcal{M}$ is large, the transition layer of $\phi$ is well-preserved. However, this results in the rounding of corners, as seen in the rectangular box. In addition, the contact zone's accuracy may be compromised, as demonstrated in \cref{subsec:hertz_rigid}.
Conversely, when $\mathcal{M}$ is small, each step of the analysis introduces significant distortions in the shape of $\phi$.
Hence, choosing an adequate value $\mathcal{M}$ is crucial to a good result.

\subsection{Contact between a growing body and a complex domain}
\label{subsec:butterfly}

\begin{figure}[tb]
    \centering
    \includegraphics{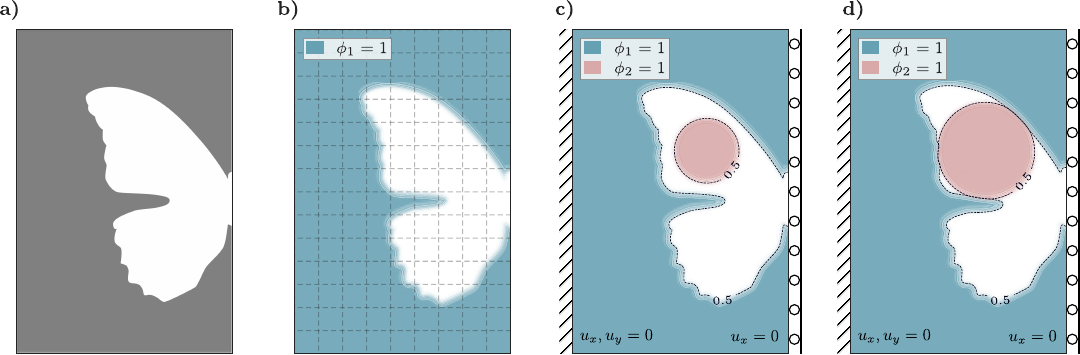}
    \caption{Simulation setup and workflow of the butterfly example.
        \textbf{(a)} Binary image of the complex domain.
        \textbf{(b)} Initial phase-field employing the hyperbolic tangent shape based on the signed distance computed from the image in (a). In addition, a schematic of the structured mesh is shown.
        \textbf{(c)} Initial configuration of the simulation. The left boundary is fixed for the blue phase, and we employ a symmetry boundary condition at the right boundary.
        \textbf{(d)} Configuration at a later step in the analysis when the disk has grown, introducing contact by which both bodies are deformed.
        }
    \label{fig:example_butterfly_schematic}
\end{figure}

For the final application of the proposed methodology, we consider the growth of a deformable body within a deformable complex domain. 
We showcase the ease of dealing with geometrically complex domains. By leveraging an Eulerian approach, representing complex geometries becomes straightforward. The structured mesh, combined with the phase-field representation and our proposed approach to handle contact, simplifies the handling of complex geometries and expands the versatility of our method for applications involving intricate domains. To illustrate this, we consider the example of a butterfly-shaped subdomain, in which we will grow a circular disk.

\Cref{fig:example_butterfly_schematic}{a} shows the binary image used for this example. To incorporate this butterfly shape with a diffuse interface into our structured mesh, we utilize the fast marching method~\cite{sethian_level_1999} to compute the signed distance function from the sharp interface. This allows us to generate the phase-field by employing the hyperbolic tangent shape given by \cref{eq:interface_profile}. 
\Cref{fig:example_butterfly_schematic}{b} illustrates the initial phase-field for the butterfly on a schematic representation of the grid.

Starting from the configuration shown in \cref{fig:example_butterfly_schematic}{c}, we will grow the disk to induce deformation, contact, and ultimately stress in this system. We grow the disk at every time step by advecting the phase-field $\phi_{2}$ with a constant displacement normal to the interface, \ie $\phi_{\text{new}} = \phi_{\text{old}} - 0.01 (\vector{n}\cdot\nabla\phi_{\text{old}})$.  Although, for simplicity, we chose an artificial growth law, which does not represent any physical process, one could easily employ a physically consistent law to grow a body.
We fix the displacement of the blue field on the left boundary and employ a symmetry boundary condition on the right boundary.
\cref{fig:example_butterfly_schematic}{d} shows the configuration in a later computational step, where the two bodies have come into contact, and the butterfly-shaped body is deformed. The $\phi=0.5$ level-sets are practically in perfect contact, even though there is always a small overlap to allow a contact zone.

\begin{figure}[tb]
    \centering
    \includegraphics{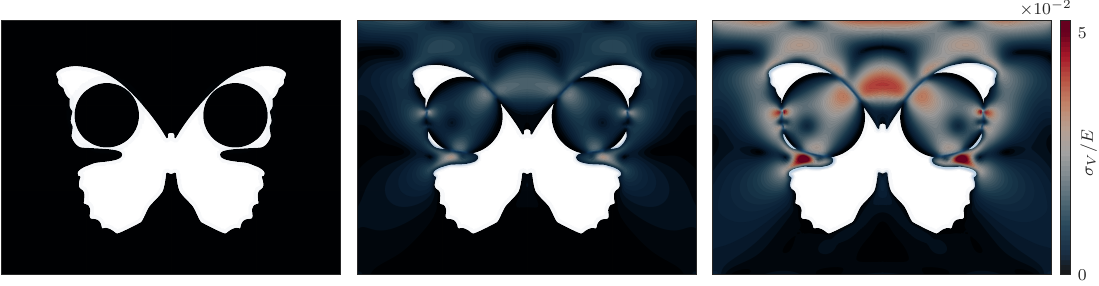}
    \caption{Von Mises stresses for three distinct steps of the butterfly example. From the initial configuration, the disk is grown to introduce contact and ultimately stress, showcasing the applicability of our method to geometrically more intricate domains.
    Evolution from left to right, from a stress-free to a stressed system. White and black color indicates void ($\phi<0.5$) and solid ($\phi>0.5$) space, respectively. The stress data is filtered and only displayed in the solid phase.
    }
    \label{fig:example_butterfly}
\end{figure}

In \cref{fig:example_butterfly}, the Von Mises stresses are shown for three steps of the analysis. The data is filtered to display only the stress where the respective phase-field is above $0.5$. Solid and void space are colored in black and white, respectively. The largest stress arises in the cantilever type overhang of the butterfly. 

The example demonstrates that dealing with more complex domains requires no additional effort. We can effortlessly incorporate intricate shapes into the computational domain by generating the initial phase-field from a binary image representation of the complex geometry. In a possible application to model the growth of precipitates in a pore network, such a binary image may come from computer tomography of the porous matrix.

\section{Discussion}
\label{sec:discussion}

The multi-field Eulerian approach using the phase-field method has the potential to be a powerful and accurate tool for modeling contact between elastic solids. Its key advantage lies in the significant reduction of system complexity. The setup of a system with multiple bodies only requires binary images of each body. These bodies can have any intricate geometrical shape without increasing model complexity.
The natural regularization of the phase-field by considering the Cahn-Hilliard equation avoids reinitialization steps often required in the context of the level-set method~\cite{valkov_eulerian_2015}.
An inherent advantage of the Eulerian framework is its natural consideration of geometrical nonlinearities. By always considering equilibrium in the deformed configuration, this approach can accurately capture large deformations and shape changes. These features make the multi-field Eulerian framework suitable for modeling complex mechanical responses, contact interactions, and intricate geometries.

The model requires careful selection of three numerical parameters: $\mathcal{M}$, $\epsilon$ and $k$. The values of $\epsilon$ and $k$ need to be chosen in accordance with the mesh resolution, while the adequate value for $\mathcal{M}$ is mainly influenced by the expected advection step size. 
The choice of the mobility parameter $\mathcal{M}$ significantly influences the accuracy and behavior of the model. A large $\mathcal{M}$ naturally regularizes the hyperbolic tangent profile and, thus, ensures that the diffuse interface is not overly distorted during its advection. However, this can introduce undesired displacements due to the gradient flow minimizing the free energy. This is particularly visible in a rounding of corners.
On the other hand, a low $\mathcal{M}$ can result in excessive distortion of the phase-field profile due to its advection. Hence, selecting an appropriate value for $\mathcal{M}$ is crucial to strike a balance between preserving the diffuse interface and minimizing the elastic strain energy.
However, based on our numerical examples, we observed that the sensitivity to $\mathcal{M}$ is relatively low in the sense that values within a few orders of magnitude usually produce satisfactory results.
The effect of the mobility coefficient $\mathcal{M}$ for interface capturing and the desire to minimize $\mathcal{M}$ is well known in literature on two-phase flow and has been discussed, $e.g.$, in~\cite{aland_phase_2017,reder_phase-field_2021,joshi_positivity_2018,dadvand_advected_2021,mao_variational_2021}.
While we have considered a constant mobility for brevity, more sophisticated approaches have been proposed.
In an attempt to dynamically determine the minimal required value, Mao et al.~\cite{mao_variational_2021} have shown an interface preserving adjustment where they introduce a time-dependent mobility $\mathcal{M}(t)$ according to the normal velocity gradient in the interface's normal direction. A more radical approach was presented recently where they propose an additional, gradient-minimizing velocity field, which is designed to eliminate variation in the advective velocity normal to the diffuse interface~\cite{mao_interface_2023}. 
Furthermore, due to the penalty approach used in our contact implementation, some level of penetration between bodies in contact is inevitable. To obtain accurate results, it is crucial to meticulously select the mesh size along with appropriate values for $\epsilon$ and the penalty constant $k$. Ideally, the intersection area should consist of multiple layers of elements to enable a smooth force distribution within the contact zone.

The framework's simplicity comes at the cost of increased system size, as each considered body introduces $2\mathrm{D}+2$ degrees of freedom per node to the solution space.
The advection required in an Eulerian framework further imposes limitations on the step size. The trade-off between the benefits of simplicity and computational efficiency depends on the specific use case and will be investigated in future studies.

In pursuit of a holistic framework, the proposed methodology can naturally accommodate surface evolution, allowing for the growth of objects within the same computational framework. By incorporating additional source terms in the flux of the phase-field, the evolution of $\phi$ can be extended to enable multiphysics coupling of contact between chemically evolving bodies.
The coupling of the proposed method with growing solids subject to solidification is the subject of our future work in line with~\cite{pundir_fft-based_2023}. 

Looking forward, our methodology can be extended to include frictional contact, given the readily available slip information from the difference of multiple displacement fields. Adhesion could be considered by expanding the intersection variable $\theta$ to encompass the negative domain in specific cases.
Furthermore, introducing Lagrange multipliers to control the contact force amplitude could improve robustness.
Lastly, further investigations are necessary on the interplay of the numerical parameters and the mesh resolution.

\section{Conclusion}
\label{sec:conclusion}

In this paper, we presented a novel Eulerian method for modeling solid-solid contact, incorporating the phase-field method. The model is based on two body-specific field variables. The phase-field $\phi$ represents the domain and interface, and its evolution is governed by an advective Cahn-Hilliard equation.
The reference map, advecting the reference configuration, is used to model elastic material.
Exploiting the intersection of the phase-fields of multiple bodies, a contact penalty force field is constructed, which is combined with the linear momentum equation to solve for the statically equilibrated configuration of the system. By employing multiple displacement fields, the solids are rendered independent, only coupled through the body forces emerging from contact between them.

We provided several numerical examples. We showed the method's accuracy for small strains by validation against the analytical Hertz solution for infinitesimal contact. 
The analysis of a disk compressing a box proved the method's applicability for problems involving large deformation, including geometrical and material non-linearities. 
In our last example, we showcased the simplicity of including evolving surfaces by resolving contact constraints between more 
complex geometrical domains with minimal additional effort.
With our numerical examples, we also showed the effect the diffuse interface width and the Cahn-Hilliard mobility coefficient have on the characteristics of the solution.

While in this paper, we have restricted simulations to contain two solids, the framework allows the inclusion of any number of objects.
By using the phase-field method for domain representation, we believe the presented model holds great promise for addressing a wide range of practical applications, particularly in the field of deterioration mechanisms where growing precipitates in a porous media induce stress and damage.

\section*{Acknowledgments}
DSK and MP acknowledge support from the Swiss National Science Foundation under the SNSF starting grant (TMSGI2\_211655).

\section*{Data Availability}
The code and generated simulation data from this study have been deposited in the ETH Research Collection database under accession code \href{}{[url will be inserted during the proof process]}.

\bibliography{references}
\end{document}